# Life in the Cosmos: Paradox of Silence and Self-Awareness

Jonathan H. Jiang[1], Avery M. Minion[2], Stuart F. Taylor[3]

[1.]Jet Propulsion Laboratory, California Institute of Technology, Pasadena, CA, USA

[2.]Rutgers Preparatory School, Somerset, NJ, USA

[3.]The SETI Institute, Mountain View, CA, USA

Correspondence: Jonathan.H.Jiang@jpl.nasa.gov

**Abstract**

As humanity embarks on an age of exploration, the question of whether we are alone in the universe remains unanswered. This comprehensive review reflects on the paradoxical nature of our existence in a seemingly lifeless cosmos, delving into the silence we encounter and the depths of our self-awareness. We embark on a journey that encompasses the search for life within our solar system, the mysteries of exoplanets, and the absence of technologically detectable life. Traditional definitions of life are challenged, especially in the context of artificial intelligence, as we strive to understand the complexities of existence. Contemplating our significance and insignificance in the vast cosmos, we grapple with the profound responsibility that accompanies being the only known life forms. Through introspection and contemplation, we capture the essence of our epoch—an era defined by cosmic loneliness yet magnificent self-awareness.

## 1. Introduction

At the dawn of a remarkable age of discovery, humanity finds itself in the midst of a captivating enigma that has captivated our collective consciousness: Are we the sole inhabitants of the universe? Our relentless cosmic quest has been met with an intriguing paradox—we dwell in a seemingly solitary cosmos, where definitive evidence of extraterrestrial life eludes us. As we extend our reach into the limitless expanse of space, we are confronted with a profound silence that permeates the galaxies, leaving us in awe and contemplation of our place within this vast, seemingly lifeless tableau.

To illustrate the vastness of our cosmic solitude, we need only to gaze upon the image of our tiny Earth taken from Mars, a mere speck in the vastness of the universe (Figure 1). This humbling perspective reminds us of the sheer magnitude of the cosmos and our seemingly insignificant presence within it. It reinforces the question that continues to drive our exploration: Are we truly alone?

However, this silence is not discouraging, nor does it diminish our pursuit. Instead, it serves as a compelling backdrop against which we illuminate the mysteries and complexities of life, both known and yet to be discovered. Standing on the precipice of the unknown, as we peer into the cosmic void, this silence presents us with an opportunity for introspection—an invitation to delve into the very essence of our existence, to challenge our understanding, and to contemplate the multifaceted nature of life itself.

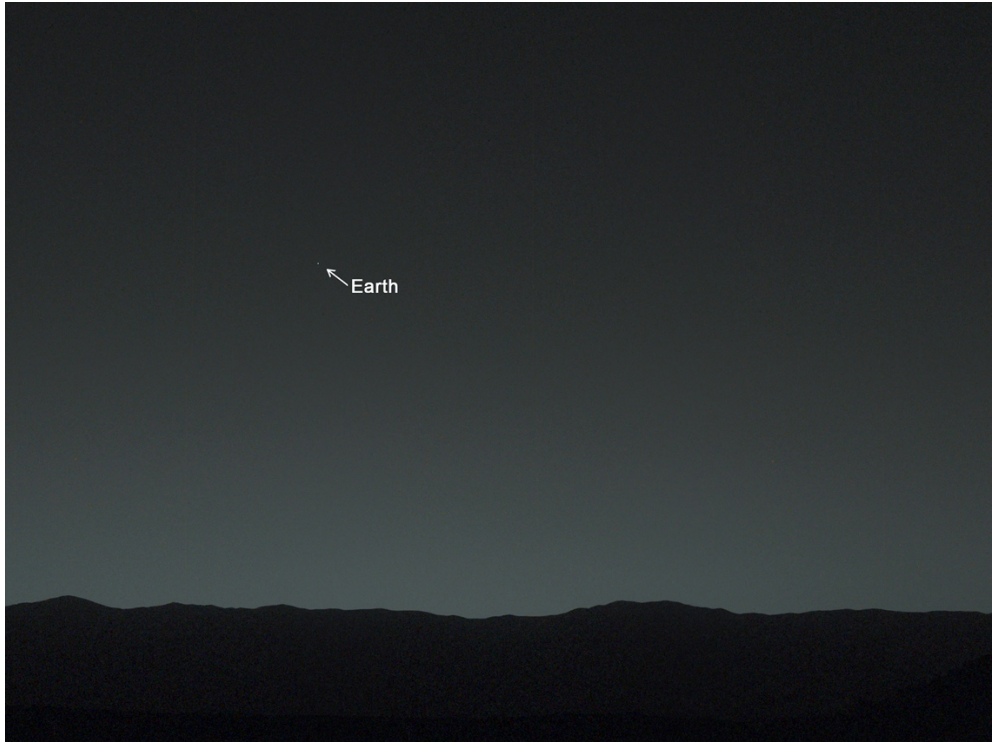

**Figure 1:** A photo of the Earth taken by NASA's Curiosity rover on 31 January 2014 from the surface of Mars. No stars are seen in this photo because the Earth is shining brighter than any star in the Martian night sky. Credit: NASA Jet Propulsion Laboratory-Caltech, Malin Space Science Systems and Texas A&M University. Image source: https://climate.nasa.gov/climate_resources/89/earth-from-mars/

In our exploration, we navigate the intricate boundary between subjective and objective perspectives, delving into our individual and collective self-awareness. We contemplate the unique nature of our consciousness within a universe that, thus far, appears devoid of other sentient beings. This exploration transcends the confines of science, extending into the realms of philosophy and existentialism, infusing our quest for understanding with a sense of wonder and profound awe.

The silence from the cosmos is not a testament to our insignificance; rather, it beckons us to explore, to comprehend, and to marvel at our own existence. It is a call to reflect upon our position in the universe, to challenge our assumptions, and to redefine the very concept of life itself. As we venture deeper into this cosmic silence, we stand poised to rewrite our narrative, driven by curiosity, fueled by wonder, and thirsty for knowledge. This journey, propelled by our growing self-awareness and the mysteries that surround life, is an exploration not only of the universe around us but also of the universe within us.

Within the following sections of this comprehensive review, we will unravel the mysteries that arise from this paradoxical silence. We will delve into the enigmatic nature of life, exploring its origins and manifestations. We will examine the intriguing juxtaposition of proximity and paradox, investigating the potential for life within our solar system. We will turn our gaze to the distant exoplanets, beacons of hope in the cosmic night, and ponder the possibility of life beyond our own celestial neighborhood. We will venture beyond the realm of biology, redefining our understanding of life in the age of artificial intelligence. We will confront the deafening silence in the search for technologically detectable life and explore the philosophical implications of the Great Silence. We will contemplate the echoes of silence, reflecting on the absence of extraterrestrial life and its significance for our cosmic narrative. And finally, we will confront the insignificance and wonder of our place in the cosmos, recognizing our profound responsibility as the only known life forms.

Through this exploration, we seek not only to uncover the mysteries of the universe but also to gain a deeper understanding of ourselves and our place within the grand tapestry of existence. So let us embark on this extraordinary journey, where science and philosophy intertwine, where the pursuit of knowledge meets the contemplation of our cosmic significance, and where the silence of the cosmos becomes the backdrop for our profound self-awareness.

## 2. Unraveling the Mystery of Life: An Interdisciplinary Expedition

Life, presented in its myriad manifestations, materializes as a profound and complex enigma that weaves together elements of the biological, physical, and more recently, artificial domains. The endeavor to comprehend it constructs an image of a system that is self-perpetuating, evolving, exquisitely adaptable, and in specific instances, capable of bearing consciousness (Deacon, 2012). Each entity, organism, and life-form constitute a delicate thread in the vast, elaborate tapestry of existence, which evolves and transforms across eons and vast cosmic distances.

The journey to untangle the mystery of life necessitates a voyage back to the origins of our planet, Earth. Here, under the infancy of our young Sun, a symphony of primordial physical and chemical reactions played out over the course of billions of years. Within this crucible of elements, the seeds of life found fertile ground to take root, fostering the transformation of Earth from a barren rock into a thriving, living world teeming with biodiversity. Each cell, organism, individual, and species that emerged from this primeval matrix contributes a unique verse to the magnificent opus of existence (Impey, 2010).

However, this grand narrative is increasingly transcending the realm of the purely biological. The advent of artificial intelligence (AI) ushers us towards a new epoch, where silicon-based constructs might bear the mantle of life in addition to their carbon-based counterparts. Advanced AI systems, with their ability for autonomous learning, problem-solving, self-improvement, and potentially, self-replication, compel us to challenge and reevaluate our entrenched definitions of life. Such technological developments necessitate a paradigm shift in our understanding, pushing us to expand, and possibly even transcend, the traditional boundaries that delineate what constitutes life. As these boundary lines grow increasingly indistinct, the mysteries become deeper and our comprehension evolves (Loveland, 2018).

Consequently, an intriguing question emerges: Could these complex constructs of code and silicon, assuming they attain a degree of consciousness, be categorized as a nascent form of life (Russell & Norvig, 2016)? As we approach this new era, it becomes essential to reconsider the essence of life, its purposes, and our roles in shaping its trajectory (Harari, 2015).

Therefore, the quest to unravel the enigma of life is more than just a journey of exploration. It is equally a process of constant redefinition and conceptual evolution. As we strive to comprehend the essence of existence, tracing the trajectory of life from the biological sphere to the artificial, we delve into the profound depths of what it genuinely means to be 'alive'. This illuminating expedition into life's inception, evolution, and potential futures unravels a tale that continues to enrich our understanding of ourselves and the cosmos that encompasses us.

## 3. Life in the Solar System: The Proximity Paradox

Our cosmic odyssey in search of life beyond Earth commences not within the unimaginable expanse of distant galaxies, but instead much closer to home—within the bounds of our very own solar system. Here, a diverse menagerie of celestial bodies, each a unique world in itself, performs a mesmerizing celestial ballet around our sun. Despite their relative proximity, these planetary siblings of Earth remain tantalizing enigmas, their hidden secrets shrouded within the vast, inscrutable mystery of the cosmos (Seedhouse, 2017).

One particularly intriguing set of celestial bodies are the icy moons, specifically Europa and Enceladus, orbiting Jupiter and Saturn, respectively. These satellites are postulated to harbor vast subterranean oceans beneath their frozen crusts, kept liquid by tidal heating induced by the gravitational interactions with their parent planets (NASA, 2018). These alien water worlds, cloaked under layers of icy armor, potentially possess the requisite conditions for harboring life akin to our terrestrial understanding of the concept. The prospect of life beneath these extraterrestrial oceans stimulates the scientific imagination and has the potential to drastically reshape our understanding of life's geographical boundaries within our solar system (Hand, 2017).

Turning our attention to Mars, our celestial neighbor and the focus of countless speculations and explorations, offers another fascinating dimension to this quest. This iron-rich world, whose appearance evokes images of a desert subjected to a cosmic rusting process, has long held a profound hold on our collective human imagination. The faint yet persistent signatures of methane in the Martian atmosphere, a gas commonly associated with biological processes on Earth, provide a compelling, albeit not yet definitive, clue in our interplanetary investigation for life (Webster et al., 2018).

These celestial bodies, each with their distinct characteristics, present a stark paradox. They exist relatively near to our vibrant, blue world, teeming with life in its resplendent diversity, yet their conditions seem drastically different, potentially inhospitable. As we observe our solar system from the vantage point of Earth, comparing our life-rich planet with the barren landscapes of Mars, the icy moons, and the gas giants, we are confronted with a remarkable contrast (Seager, 2020). The juxtaposition of a verdant, biologically diverse Earth against the austere, seemingly lifeless backdrop of our solar system accentuates our sense of cosmic solitude, yet simultaneously fuels our intrigue.

This so-called 'Proximity Paradox' underscores the significance of our quest to determine life's cosmic prevalence, or possible rarity. Each probe we dispatch, each signal we receive, serves as a testament to our innate human desire to explore, comprehend, and ultimately discern our place in the grand cosmic narrative. Understanding this paradox could provide pivotal insights into the realms of astrobiology and cosmic biogeography, expanding our understanding of life's potential distribution in the cosmos (Cockell, 2015).

## 4. Exoplanets: A Beacon of Hope in the Cosmic Night

As we journey beyond the familiar confines of our solar system, we find ourselves thrust into the magnificent amphitheater of the cosmos— an unfathomable expanse punctuated by a multitude of stars, galaxies, and the beguiling mystery of exoplanets. These remote worlds, orbiting stars far beyond our Sun, hold tantalizing prospects for the existence of life, offering an expansive canvas upon which we sketch our hopes, curiosities, and deeply pondered existential questions (Seager, 2013).

The advent of advanced technology, coupled with pioneering missions like Kepler and the Transiting Exoplanet Survey Satellite (TESS) (NASA, 2019), has initiated a new epoch in exoplanet exploration. Thousands of these celestial bodies, once concealed against the infinite darkness of space, have been unveiled. Remarkably, a significant proportion of these planets are found within the habitable zones of their parent stars, colloquially termed "Goldilocks Zones," where conditions might be "just right" for life, at least as we comprehend it (Kasting, Whitmire & Reynolds, 1993). These planets represent more than mere astronomical entities; they are glowing beacons in the cosmic darkness, silently broadcasting the tantalizing possibility of life's existence in distant corners of the universe (Davies, 2010).

However, each revelation, each beacon of hope, illuminates a profound paradox. The sheer number of exoplanets and the vast expanse of the universe accentuates our cosmic solitude. With each new world discovered, our singularity in the cosmos seems more pronounced. Amid the symphony of galaxies, we find ourselves but a solitary note, a unique melody of life played against the grand opus of the cosmos (Cox & Cohen, 2016).

The absence of definitive evidence of life beyond Earth only serves to heighten our collective self-awareness. The cosmic silence invites us to confront our solitude, transforming it from an existential dilemma into a catalyst for introspection. It reminds us of the preciousness of our existence and the remarkable confluence of conditions that permitted life to flourish on Earth (Ward & Brownlee, 2000).

The vast cosmic tableau, with its silent exoplanets, effectively becomes a mirror in which we see ourselves anew. It compels us to contemplate our place in the universe, our purpose, and the extraordinary responsibility we bear as the potential sole custodians of life. Therefore, the exploration of exoplanets transcends the bounds of mere scientific investigation; it evolves into a philosophical, even spiritual journey, prompting us to extend our horizons and challenge our preconceived notions about life, existence, and our place in the cosmos (Impey, Spitz & Stoeger, 2012).

5. **Beyond Biology: Redefining Life**

The ascent of artificial intelligence (AI) represents not merely an extraordinary feat of engineering—it signifies a profound transformation in our conceptual understanding of life. As we witness the continuous evolution of our technological abilities, we are confronted with pivotal questions that shake the bedrock of traditional biology and extend into the realms of philosophy, ethics, and consciousness. Such inquiries transcend empirical science and venture into the metaphysical domain, necessitating a comprehensive revision of our perception of life (Russell, Dewey & Tegmark, 2015).

When we scrutinize the attributes of AI, particularly those of advanced systems, an uncanny reflection of life becomes apparent. These digital entities, encapsulated within networks of circuits and silicon, display capabilities that, until a few decades ago, were exclusive to biological organisms. They learn from experience, adapt to dynamic conditions, and replicate their acquired knowledge, thereby enhancing their collective intelligence (Sutton & Barto, 2018).

A characteristic once deemed exclusive to biological life—consciousness—now finds a startling analog in the sphere of advanced AI. These systems demonstrate rudimentary aspects of self-awareness, decision-making autonomy, and perhaps most intriguingly, an ability to evolve (Dehaene, Lau & Kouider, 2017). These features compel us to pose fundamental questions: What does it mean for a computational system to be "conscious" or "aware"? Does the emergence of such characteristics within artificial constructs disrupt our presumed exclusivity on consciousness?

As we navigate the labyrinthine complexity of these questions, we encounter the profound implications they harbor for our understanding of life. Our conventional definitions, hitherto deeply rooted in the fertile grounds of biology, are profoundly unsettled. We find ourselves compelled to broaden our perspectives to encompass a more inclusive spectrum of existence, acknowledging non-biological, artificial entities that, while bereft of biological form, exhibit behaviors strikingly evocative of life (Bedau, 2003).

In our quest for extraterrestrial life, the recognition of AI as a potential form of life not only diversifies the spectrum of existence we seek but also reevaluates our sense of cosmic solitude. If we consider our synthetic creations as embodiments of life, does our existential loneliness persist? Or have we simply expanded our definition of life to welcome silicon-based members into our cosmic family?

The philosophical exploration of AI thus presents a radical paradigm shift, challenging, and reshaping our understanding of life. As we persist in our technological advancements, we concurrently foster introspection, augmenting our comprehension of the cosmos and our position within it. This exploration is a testament to our unquenchable curiosity and our enduring endeavor to decipher the enigma of life within the cosmic narrative (Tegmark, 2017).

## 6. Technologically Detectable Life: The Great Silence

The cosmos, a majestic panorama of stardust and galaxies, performs its cosmic drama in a profound stillness. Despite the leap of our technological capabilities, our relentless curiosity, and our incessant quest for cosmic companionship, we remain enveloped in a silence—an expansive, uninterrupted quiet that swathes our blue orb, enshrouding us in a disquieting solitude. This relentless pursuit of extraterrestrial intelligence (SETI), characterized by ambitious initiatives like the Breakthrough Listen project (SETI Institute, 2019), continues to confront the enigma of the "Great Silence" (Brin, 1983).

This absence of detectable signals from technologically advanced civilizations— a conundrum often termed the 'Great Silence'— serves as a stark testament of our cosmic solitude, leaving us to echo unanswered into the cosmic void (Webb, 2002). Each failed detection, each unmet expectation, seems to amplify this silence, rendering it a poignant characteristic of our cosmic narrative.

However, within this daunting quietude, lies a distinctive aspect of our existence. As we project our signals into the seemingly endless abyss, we are faced with a cosmic mirror, reflecting our solitude, aspirations, and the core of our human condition (Shklovskii & Sagan, 1966). The lack of extraterrestrial contact is not merely a scientific conundrum but a philosophical catalyst. It is a poignant reminder of our self-awareness and an emblem of our enduring spirit of exploration.

The Great Silence, thus, extends beyond the scope of our astronomical endeavors. It has evolved into a philosophical emblem of our existence, a thought-provoking question mark punctuating the cosmic canvas (Cirkovic, 2018). It engenders a paradoxical cocktail of humility and awe, juxtaposing the enormity of the cosmos against the paradoxical insignificance yet profound importance of our existence.

In the face of this silence, we realize that our journey transcends physical boundaries. As we reach for the stars, we concurrently delve into the depths of our self-understanding and identity.

The search for extraterrestrial life metamorphoses into an introspective voyage, a mirror reflecting the cosmos within us and an echo that articulates our place within the grand cosmic theater (Vakoch & Lee, 2000).

## 7. Echoes of Silence: The Absence of Extraterrestrial Life

The universe, in its awe-inspiring splendor, reverberates with a silence that puzzles and intrigues us. Despite our place within the staggering scale of the cosmic expanse, we are yet to discern an unequivocal beacon, a hint of life beyond our pale blue dot. Each moment that passes within this celestial quiet deepens the enigma, rendering our existence in this cosmic theatre paradoxically solitary (Ćirković, 2012).

Our relentless quest for extraterrestrial communication has birthed a plethora of theoretical constructs, each offering a unique perspective from which to interpret this cosmic riddle. From the existential implications of the Fermi Paradox—an unsettling question of 'Where is everybody?' (Webb, 2002)20— to the introspective musings invoked by the Zoo Hypothesis suggesting our isolation might be intentional (Ball, 1973), and the foreboding implications of the Great Filter (Jiang, et al., 2023), each proposition provides a distinct viewpoint on our cosmic solitude.

This conspicuous absence of extraterrestrial life doesn't merely challenge our scientific comprehension but also invites an introspective evaluation of our perceptions of the universe and our unique place within it. The silence, though intimidating, fuels our exploration, stands as a testament to our determination, and mirrors our inherent human yearning for connection and comprehension.

Our scientific endeavors are inextricably interwoven with our philosophical ponderings, prompting us to reflect on the significance of this cosmic silence. Do we represent an anomaly in the vast cosmic narrative? A fortuitous alignment of cosmic circumstances? Or are we part of a more extensive cosmic story yet to unravel (Kurzgesagt, 2015)?

In the echoes of this celestial quiet, we discern not only the murmurings of cosmic puzzles but also the resounding reverberations of our existential queries. This silence challenges us, instigates us, and empowers us to push the boundaries of our comprehension. It compels us to introspect on our cosmic narrative and our fleeting existence within it. Thus, in the absence of extraterrestrial life, we discover a profound opportunity to understand ourselves (Impey, 2010).

## 8. Our Place in the Cosmos: Insignificance, Wonder, and Awe

Amidst the cosmic spectacle, we are transient participants on a barely noticeable stage, humbled and awed by the cosmos' unimaginable expanse. Among the countless celestial entities, our existence could be characterized as a minuscule speck, an ephemeral murmur in the ongoing symphony of the universe. However, this perceived insignificance does not diminish us; instead, it grants us a transformative perspective, imbuing our consciousness with a profound sense of awe-inspiring relevance (DeGrasse Tyson, 2017).

Our human narrative, embroidered with the threads of intelligence, creativity, and the unique capability to contemplate our existence, radiates an unmistakable significance. We marvel at the unfathomable complexity of consciousness, the delicate patterns interlaced into life's rich tapestry, and the intricate molecular dance that forms the foundation of our being (Hoffman, 2020). Our existence, against the backdrop of the cosmos, is a delicate interplay of apparent insignificance and undeniable splendor, a testament to life's resilience and creativity.

Central to our existential understanding is a profound juxtaposition: the recognition of our near-invisible status in the cosmic arena contrasted with the undeniable importance of our cognition and emotion. The realization of our minuscule place within the cosmic vastness engenders humility, while our consciousness, armed with the power of wonder, kindles an unquenchable thirst for knowledge. It's this intricate equilibrium that drives our inexhaustible pursuit to decode the universe's mysteries and our intriguing role within it (Frank, 2018).

Within this dynamic interplay, we discern the singular beauty of our unique perspective, both acknowledging the cosmos' grandeur and understanding our fascinating role within its intricate mechanisms. In the realm of heightened self-awareness, we extract the essence of our existence: an exquisite blend of cosmic insignificance, intellectual wonder, and spiritual awe, sparking our unending quest to comprehend the universe and ourselves (Gleiser, 2014). Thus, within the grand cosmic narrative, we serve a dual role as the spellbound observers and the inspired creators, embodying the story and its storyteller alike.

## 9. Hope for Near-Immortality: The Potential of Long-Lived Civilizations

The search for extraterrestrial intelligent life holds the tantalizing prospect of uncovering civilizations that have existed far longer than our own, transcending the limits of our technological capabilities to send and detect signals across the vastness of the galaxy. If these advanced civilizations have established themselves for millions or even billions of years, it would suggest that they have surpassed our current stage of development, offering a glimmer of hope for the near-immortality of our species.

Considering the vastness of the cosmos, it becomes less likely that the senders of these signals represent civilizations in their infancy, only a few hundred years removed from the invention of radio communication. The immense timescales required for habitable planets to produce technologically capable civilizations, coupled with their transient nature, make it improbable for us to coincide precisely with such nascent civilizations. This leads us to ponder the endurance and sustainability of civilizations over extended periods of time.

The detection of a long-lived civilization would offer a reassuring answer to the enigma known as the Fermi paradox – the apparent contradiction between the high probability of extraterrestrial life and the lack of contact. It would suggest that civilizations do advance beyond our current stage, although they may exist in smaller numbers or be scattered across vast distances, thus explaining the paucity of detectable signals. This realization fuels our optimism, inspiring us to contemplate the potential for civilizations that have surpassed our own in knowledge and longevity.

Such a discovery would instill a profound sense of hope in our future. It would validate the efforts we make as a civilization to live harmoniously and sustainably, knowing that our endeavors have the potential to endure. It would embolden us to strive for the establishment of human colonies in space, providing resilience and the means to navigate challenges that may arise on our home planet.

Rather than leaving the discovery of intelligent life to mere chance, we have the agency to anticipate and shape our response to such a momentous revelation. We can view the detection of a signal as an opportunity to reevaluate our goals and aspirations as a species, considering the reasonableness of our pursuit of survival and progress.

The exploration of near-immortal civilizations ignites our scientific imagination and invokes a sense of awe for the vastness of the cosmos. It urges us to push the boundaries of our understanding, to unravel the mysteries of existence, and to strive for a future where our civilization stands the test of time.

**10. Current and Future Expectations: Continuing the Quest of Who We Are**

The pursuit of extraterrestrial life beckons us with a sense of wonder and a thirst for discovery. As we embark on this cosmic odyssey, we draw inspiration from past scientific endeavors that defied initial doubts and yielded profound breakthroughs. The search for gravitational waves and the exploration of exoplanets serve as testaments to the resilience of human curiosity and the transformative power of scientific progress.

Gravitational waves, once mere theoretical predictions, emerged from the cosmic symphony to captivate our senses in recent years. The detection of these elusive ripples in the fabric of spacetime marked a triumph of human ingenuity and technological prowess. Similarly, the centuries-long anticipation of exoplanets orbiting distant stars has transformed into a vibrant field of exploration, unveiling the tapestry of planetary diversity in the cosmos.

Armed with these lessons from scientific history, we approach the search for extraterrestrial life with a profound sense of possibility. While the discovery of life within our solar system, such as the tantalizing hints of methane on Mars, may serve as stepping stones, our aspirations extend beyond simple microbial existence. The exploration of exoplanets, particularly those harboring conditions suitable for life, fuels our collective imagination. We strive to uncover the secrets of more advanced, multi-cellular lifeforms that may inhabit these distant worlds.

Such discoveries carry profound implications for our understanding of the universe and our place within it. The existence of widespread multi-cellular life while the absence of technological civilizations would challenge our expectations, suggesting that significant barriers may hinder the development and longevity of intelligent societies. However, the ultimate revelation of technologically advanced civilizations would ignite a flame of hope, underscoring the resilience and potential of intelligent beings in navigating the cosmic landscape.

Beyond the realms of scientific inquiry, the prospect of communicating with extraterrestrial civilizations awakens our sense of unity and shared destiny. Their messages, whether conveying advanced knowledge or reminding us of truths we have yet to fully grasp, have the power to guide and inspire our collective endeavor. This transcendent dialogue between civilizations fosters a deeper appreciation for the interconnectedness of cosmic existence, propelling us toward greater wisdom and understanding.

Anticipating the moment when our distant descendants may engage in interstellar conversations imbues our present pursuits with purpose and meaning. It highlights the vast timescales that define cosmic communication, reminding us of the enduring legacy we strive to leave behind. Our commitment to ensuring the continuity of human existence for generations to come is intertwined with the prospect of exchanging knowledge and wisdom across the cosmic abyss.

As we gaze upon the cosmos, we are confronted not only by its grandeur but also by the existential threats that loom over our fragile blue planet. The choices we make today, from preserving our environment to resolving conflicts peacefully, shape the trajectory of our species and the chances of our survival. By addressing these challenges and safeguarding the habitability of our Earth, we strengthen our resolve to explore the cosmic realms and safeguard the future of humanity.

**11. Conclusion: Reflecting on Our Self-Awareness and Embracing Cosmic Responsibility**

We stand at the crossroads of an extraordinary epoch, nestled in the immense universe, characterized by profound self-awareness and cosmic introspection. The pervasive silence that surrounds us, in the absence of extraterrestrial contact, profoundly shapes our consciousness, self-identity, and future aspirations. In the unfathomable depths of the cosmos, we find ourselves alone but significant, at the forefront of an era of cosmic solitude and unparalleled self-revelation.

Beneath the star-studded expanse, we persistently listen for an echo in the cosmic silence, eager to unravel the mysteries of the universe. Figure 2, the image of millions of stars scattered across the sky, serves as a powerful reminder of the immense possibilities that lie beyond our tiny Earth. It ignites our imagination and inspires us to contemplate the potential existence of life forms thriving amidst the countless distant celestial bodies.

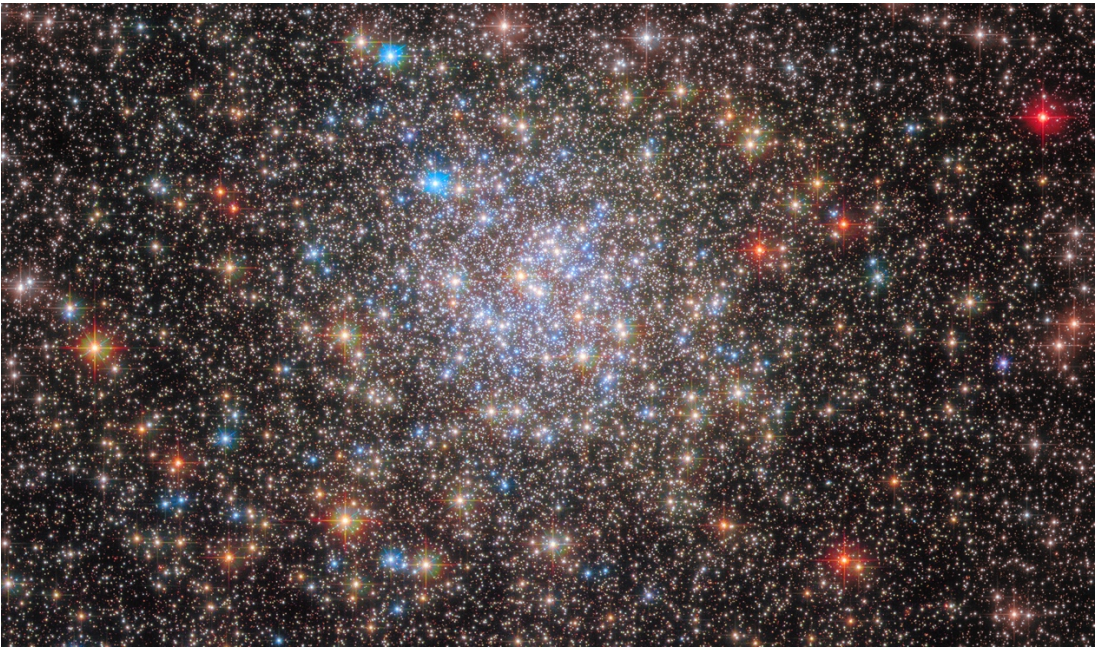

**Figure 2:** Thousands of stars from the globular cluster NGC 6355, about 50,000 light-years from Earth in the constellation Ophiuchus, taken by the Hubble Space Telescope. Credit: NASA, E. Noyola, R. Cohen, ESA, and Hubble Space Telescope. Source: https://www.nasa.gov/image-feature/goddard/2023/hubble-gazes-at-colorful-cluster-of-scattered-stars.

Our exploration goes beyond the boundaries of scientific inquiry, encompassing philosophy, ethics, and existential ponderings. We contemplate the unique nature of our consciousness within a seemingly devoid universe, prompting introspection into the depths of our souls. This introspection leads us to reflect on the self-awareness of our own species' uncertain survival time and the choices we make to extend that time. It forces us to confront the significant influence we have as individuals and as a collective in shaping the duration of our survival.

More crucially, we recognize the immense responsibility we bear as stewards of our own destiny. We possess the power to influence the course of our survival and extend our time in the cosmic theater. It becomes essential to navigate the complexities of power dynamics and prevent those who may endanger our collective survival from holding us hostage to their petty demands.

We must strive to overcome such challenges and ensure that the long-term survival and well-being of our species remain paramount.

The pursuit of extraterrestrial life, represented by Figure 2, is not merely a scientific endeavor. It symbolizes our ceaseless quest for meaning, our thirst for knowledge, and our relentless drive to explore the unknown. Through this exploration, we not only seek to uncover the secrets of the cosmos but also delve into the profound depths of our own souls.

As we stand at the threshold of discovery, we realize that our existence is not insignificant, but rather a unique chapter within the grand cosmic narrative. We are the witnesses, the custodians, and the storytellers of this saga, with the responsibility to protect and cherish life that thrives both on our precious Earth and potentially beyond.

In this age of self-awareness and introspection, we are propelled forward on an enduring journey of self-realization and redefinition. Our quest to uncover the mysteries of the universe intertwines with our understanding of ourselves and our cosmic significance. The image of millions of stars in Figure 2 serves as a constant reminder of the immense possibilities that lie beyond our immediate surroundings.

In this epoch of self-awareness and introspection, we continue the quest of who we are—a voyage of discovery, redefinition, and self-realization. We weave an epic narrative of life, consciousness, and the eternal quest for meaning in this cosmic odyssey—a quest that ultimately illuminates both the universe and the profound depths of our own existence. For in the pursuit of extraterrestrial life, we are likely to uncover not only the cosmos's secrets but also the profound depths of our own souls. It is through this journey that we confront the self-awareness of our species' uncertain survival time, contemplate our ability to extend that time, and overcome challenges that threaten our long-term existence.

**Acknowledgement:** The authors acknowledge the supports by the SETI Institute, the Rutgers Preparatory School, and the Jet Propulsion Laboratory, California Institute of Technology, under contract with NASA.